# Reversible shear thickening at low shear rates of electrorheological fluids under electric fields

Yu Tian,[1]* Minliang Zhang,[1] Jile Jiang,[1] Noshir Pesika,[2] Hongbo Zeng,[3] Jacob Israelachvili,[4] Yonggang Meng,[1] Shizhu Wen[1]

1. State Key Laboratory of Tribology, Department of Precision Instruments, Tsinghua University, Beijing 100084, P. R. China
2. Department of Chemical and Biomolecular Engineering, Tulane University, LA 70118, USA
3. Department of Chemical and Materials Engineering, University of Alberta, AB, T6G 2V4, Canada
4. Department of Chemical Engineering, Materials Research Laboratory, University of California, Santa Barbara, California 93106, USA

## Abstract

Shear thickening is a phenomenon of significant viscosity increase of colloidal suspensions. While electrorheological (ER) fluids can be turned into a solid-like material by applying an electric field, their shear strength is widely represented by the attractive electrostatic interaction between ER particles. By shearing ER fluids between two concentric cylinders, we show a reversible shear thickening of ER fluids above a low critical shear rate ($<1$ s$^{-1}$) and a high critical electric field strength ($>100$ V/mm), which could be characterized by a modified Mason number. Shear thickening and electrostatic particle interaction-induced inter-particle friction forces is considered to be the real origin of the high shear strength of ER fluids, while the applied electric field controls the extent of shear thickening. The electric field-controlled reversible shear thickening has implications for high-performance ER/magnetorheological (MR) fluid design, clutch fluids with high friction forces triggered by applying local electric field, other field-responsive materials and intelligent systems.

*Correspondence to: tianyu@mail.tsinghua.edu.cn

# 1. Introduction

Colloidal suspensions have attracted much interest because of their wide range of applications, such as in decorative or protective paints, laser printer ink, and advanced pharmaceutical drug delivery.[1-3] Shear thickening is a phenomenon of significant viscosity change of colloidal suspensions and has recently been exploited to fabricate advanced flexible liquid armors.[4-6] ER and MR fluids are electric and magnetic field responsive colloidal suspensions,[7-15] the rheological properties of which can be abruptly altered from a Newtonian fluid with $\tau = \eta_0 \dot{\gamma}$ into a Bingham fluid with $\tau = \tau_E + \eta_0 \dot{\gamma}$, where $\tau$ is the shear stress, $\eta_0$ is the viscosity of the suspension under zero electric field, $\dot{\gamma}$ is the shear rate, and $\tau_E$ is the shear yield stress. Since the invention of ER and MR fluids by Winslow [7] and Robinnow, [11] shear yield stress has been represented by the electrostatic/magnetic attractive strength between ER/MR particles with various polarization models. [7-14]

A general ER fluid has a shear yield stress of several kPa and requires a local electric field between particles in the order of 10 kV/mm. [7-9,15] However, a recently developed "giant ER fluid" has shown a yield stress over 100 kPa [10,16] and was claimed to have a different mechanism from the traditional ER effect since it requires a rather high local electric field between particles in the order of 100 kV/mm. This local electric field is even higher than the breakdown electric field of most insulating liquids. The magnitudes of the local electric field and the attractive electrostatic field strength between ER particles with nanometer gap distances have not been experimentally quantified due to experimental difficulties. Experimental and theoretical studies have only been done to explore the formation of particle chains and columnar structures in the static equilibrium state, [17-19] as well as the shear-induced striped/lamellar structures, both in ER and MR fluids. [20-23] Figure 1 shows a schematic of ER particles under conditions of zero electric field, high electric fields, and high electric fields in addition to shearing. However, the relationships between the rheological properties of ER/MR fluids and their chain structure under an electric/magnetic field have not been well understood. Most available models predict the shear strength of ER/MR fluids by considering attractive electrostatic/magnetic forces due to the dipole-dipole or multipole interactions between particles, without including the structural evolution of ER/MR fluids during shearing. [8,9,13,14]

In this study, ER fluids were sheared between two concentric cylinders on a commercial rheometer (Physica MCR 301). A reversible shear thickening of ER fluids above a low critical shear rate (<1 s$^{-1}$) and a high critical electric field strength (>100 V/mm) has been found. The electric field controlled shear thickening was verified with various experimental test modes or methods, ER fluids with different particle types, particles volume fractions, insulating medium viscosities, and characterized by a modified Mason number or more generally, a critical bulk viscosity of the ER fluid under electric fields. The shear thickening and electrostatic particle interaction-induced friction force between particles is considered to be the real origin of the high shear strength of ER fluids. While the applied electric field controls the extent of shear thickening.

## 2. Materials and Methods

In this study, the ER fluid consisted of NaY zeolite particles[15] (Qilu Petro. Corp., Shandong,

China) with density of 1.85 g/cm$^3$, average diameter of 1 μm, and silicone oil (Beijing Chem. Corp., China) with viscosity of 10 mPa.s at room temperature 20 °C, density of 0.93 g/cm$^3$, and dielectric constant of 2.56. The zeolite particles were washed with de-ionized water several times and then dried in a microwave oven. A mixture of 1:9 glycerin/ethanol was then combined with the dry zeolite particle to obtain a weight ratio of 1% glycerin to zeolite particles, which results in a thin coating of glycerin on the zeolite particles. Silicone oil was used as received. Zeolite particles were mixed with silicone oil in a four-roll miller to obtain a uniform ER suspension. The shear tests were done on a commercial rheometer (Physica MCR 301, Anton Paar, Germany) that have two concentric cylinders, with an inner diameter of 16.66 mm, length of 25 mm, and gap of $h$=0.7 mm. The electric field and the shear rate are controlled by the rheometer. In the shear rate ramp tests, the shear rate is logarithmically ramped up from 0.01 to 50 s$^{-1}$ in 180 s, and then subsequently logarithmically ramped down from 50 to 0.01 s$^{-1}$ in 180 s.

## 3. Results and Discussion

### 3.1 Shear thickening of ER fluids found in shearing tests

As shown in Fig. 2a, a typical Newtonian behavior with $\eta_0$=0.06 Pa.s under zero electric field and a Bingham fluid behavior with a finite $\tau_E$ under high electric fields were obtained, as expected (as $\dot{\gamma} \to 0$, the shear stress approaches the static yield stress $\tau_s$). The results for the low shear rate region are replotted in Fig. 2b, which shows an abrupt, significant, and reversible change of shear stress, occurring around a certain critical shear rate $\dot{\gamma}_c$ and above a critical electric field $E_c$. An abrupt change of shear stress or viscosity in colloid suspensions is usually referred to as shear thickening,[24, 25] which is induced by arresting and pushing particles against one another. The occurrence of the abrupt and reversible shear stress observed in ER fluids can also be considered as shear thickening. Figure 2b shows that no shear thickening is observed when $E < 900$ V/mm. As $E$ increases from 900 to 2900 V/mm, the critical shear rate $\dot{\gamma}_c$ for shear thickening increases from 0.02 to 0.2 s$^{-1}$, and the ratio of shear stress change, $\tau_H/\tau_L$ increases from about 1.5 to 3.2 (shown in Fig. 2c).

Prior studies have shown that a shear yield strain $\gamma_y$ of about 0.3 is needed to reach the static yield stress $\tau_s$ of an ER fluid.[8,14] In Fig. 2, the shear rate is logarithmically ramped up from 0.01 to 50 s$^{-1}$ in 180 seconds and 60 points, and subsequently ramped down. Each shear rate only lasted for 3 seconds. The shearing at each shear rate has not exceeded the shear yield strain to reach a stationary state. So the shearing time effect on the result has been studied. But all the obtained results (Fig. 1 of the supporting information.) show no obvious difference from each other on the shear stress jump phenomenon around a critical shear rate. Further, to confirm that the shear thickening shown in Fig. 1b is just governed by the shear rate rather than the shear strain, experiments under the same $E = 2150$ V/mm but different constant shear rates $\dot{\gamma}$ in the range of 0.02-0.09 s$^{-1}$ were performed. The results are shown in Fig. 3a and 3b. When $\dot{\gamma} < \dot{\gamma}_c$, the shear stress remains low and independent of shear strain. On the other hand, when $\dot{\gamma} \geq \dot{\gamma}_c$, the shear stress is high even at a small shear strain of 0.1 (10%), which is less than the general shear yield strain of 0.3 (30%). The amplitudes of shear stress achieved at a different $\dot{\gamma}$ agree well with the shear rate ramp test. The critical shear rate $\dot{\gamma}_c$ between 0.085 and 0.09 s$^{-1}$ is also consistent with that of 0.08-0.12 s$^{-1}$ obtained in the shear rate ramp test. This experiment verified that the shear thickening is indeed governed by the shear rate and not the shear strain, and it also indicates that the shearing time does not affect the shear stress jump phenomenon shown in Fig. 2.

Since in this study a higher $E$ applied on ER fluid induces a larger local electric field $E_{loc}$ between ER particles, a larger bulk electrical current $I$, and a higher shear yield stress $\tau_E$,[26] the current $I$ was simultaneously measured to represent the $E_{loc}$ change during shearing. A typical result is shown in Fig. 3c. Accompanying the abrupt increase in shear stress during shear thickening, $I$ decreased sharply but only slightly. This indicates a small decrease in $E_{loc}$ between particles, which could be ascribed to the dilatancy of ER fluid that usually happens during the shear thickening or jamming of colloidal suspensions.[4-5] Dilatancy that caused less dense particle packing in an ER fluid usually corresponds to a lower $E_{loc}$.[22] During the reverse process (that is, decreasing the shear rate), $I$ slightly increased, which corresponds to the un-shear thickening process of ER fluid. This behavior is shown in Fig. 3c.

## 3.2 Mechanism of Shear thickening happened in ER fluids

Shear thickening in a colloid suspension would occur when the compressive hydrodynamic force between two particles is larger than the total repulsive force between these particles.[4,5,24,25] A scaling theory takes the ratio of hydrodynamic force to the Brownian force $\tau_{cr}^{Br} = F_{hydrodynamic} / F_{Brownian}$ as an effective dimensionless critical shear stress to predict the onset of shear thickening, which agrees well with the tests of suspensions with different particle sizes and particle concentrations.[5]

In research on ER effect, the Mason number [14]

$$Mn = \frac{\eta_c \dot{\gamma}}{2\varepsilon_0 \varepsilon_c \beta E_0^2} \quad (1),$$

which is the ratio of the viscous force to the attractive electrostatic force between the particles, has been used to characterize the rheological change of ER fluids. In Equation (1), $\eta_c$ is the viscosity of the continuous fluid, $\varepsilon_0$ is the dielectric permittivity of vacuum, $\varepsilon_c$ is the relative dielectric constant of the continuous fluid, $\beta$ is the polarization ratio of particles under an external electric $E_0$. However, the $Mn$ for the results shown in Fig. 2 does not indicate a single characteristic value for shear thickening. This might be caused by the neglect of other forces between particles. We therefore propose a modified Mason number as

$$Mn^* = \eta_c \dot{\gamma} / \tau \quad (2),$$

as the ratio of hydrodynamic stress to shear stress (including all inter-particle interactions such as electrostatic, van der Waals, hydrolubrication, friction, and repulsive steric forces). The viscosity of silicone oil in this study is $\eta_c = 10$ mPa.s. Three $Mn^*$ curves for the tests shown in Fig. 2 are plotted in Fig. 4a. The critical Mason numbers $Mn_c^* = \eta_c \dot{\gamma}_c / \tau_c$ at the onset of shear thickening (circular dots, $\tau_c$ is the critical shear stress) agree reasonably well with one another. ER fluids with different particle volume fractions (5-28%) and under different electric fields (0-5 kV/mm) also show similar $Mn_c^* = 1.2 \times 10^{-6}$ for the onset of shear thickening, and $Mn_c^* = 0.8 \times 10^{-6}$ for the onset of un-thickening. This result indicates that the shear thickening in ER fluid occurs when $Mn^*$ increases beyond $Mn_c^* = 1.2 \times 10^{-6}$, $E > E_c$, and $\dot{\gamma} \geq \dot{\gamma}_c$.

Figure 4b shows tests under different constant shear rates and ramped voltages. The $Mn_c^*$ for shear thickening and un-thickening are about $1.3\times10^{-6}$ and $0.8\times10^{-6}$, respectively (shown in Fig. 4c), consistent with the shear rate ramp tests shown in Fig. 2. Figure 4b also shows that $\tau$ increases with $E$ with different slopes $K$: $K_1$=1.1 Pa.mm/V is under the shear thickened state ($\tau = K \cdot E$, $\tau$ in Pa, $E$ in V/mm). $K_2 = 0.26$ Pa.mm/V is under the un-shear thickened state, and decreases with increasing $\dot{\gamma}$ (under a high $\dot{\gamma}$, $K_2 \to 0$). The linear slopes of $K_1$ and $K_2$ are different from $\tau \propto E^2$ predicted by the traditional polarization models.[8,9,14,26]

At this stage, shear thickening has been found for the same ER fluid tested with different shear rate control modes and different electric field applying types, and different ER materials (prepared by ourselves and donated from other institutions, as shown in Fig. 1 of the supporting information). To verify this phenomenon in more general cases, experiments with different particle volume fractions have been tested on Physica MCR 301. They all show shear stress jump phenomenon above some critical shear rates and critical electric fields, but with a similar $Mn_c^*$. Results are shown in Fig. 2 of the supporting information. $E_c$ for shear thickening to happen decreases from 4667 V/mm (4.3 %) to 533 V/mm (28 %). The value of $\tau_H/\tau_L$ increases with the increase of $\phi$ and $E$. It is above 3 for $\phi$=28 % and $E$= 2900 V/mm. Taking $\tau_H$ as the shear yield stress, the relationship between $\tau_H$ and $E$ roughly agrees with the square relationship predicted by traditional polarization models.

According to Equation 2, the insulating liquid viscosity may change the modified Mason number. So it is important to know the effect of insulating liquid viscosity on the critical modified Mason number for the shear thickening. Therefore ER fluids with the same zeolite ER particles but different medium viscosities ranging from 10 to 1000 mPa.s have been tested. (Original shearing curves are shown in Fig. 3 of the supporting information.) The obtained $Mn_c{}^*$ for the onset of shear thickening shows a great difference for ER fluids with different insulating liquid viscosities as shown in Fig. 5 (a). It is about $10^{-4}$ for the ER fluids prepared with silicone oil viscosity of 1000 mPa.s, while is about $10^{-6}$ with 10 mPa.s. The later value agrees with the results shown in Fig. 2-4. It indicates that $Mn_c{}^*$ is not general for predicting the shear thickening of ER fluids prepared with different medium viscosities. However, plotting the apparent viscosity of ER fluid versus shear rate, a critical apparent viscosity of ER fluids under electric fields for the shear thickening could be found as shown in Fig. 5(b). From Equation 2, the viscosity can be represented by the modified Mason's number as

$$\eta_{ST} = \tau / \dot{\gamma} = \eta_c / M_n{}^* \quad (3).$$

All the above experiments with a medium viscosity of 10 or 50 mPa.s, $Mn_c{}^*$ obtained could give a reasonable agreement with the critical viscosity of ER fluids under electric fields. As shown in Fig. 5(c), the shear thickening happens at about $\eta_{ST}= 0.8 \sim 2 \times 10^4$ Pa.s (also agree with results shown in the supporting information, Fig. 1 (b) and (d) and Fig. 2). However, the physical meaning of the critical bulk viscosity should be disclosed later.

## 3.3 Shearing ER fluids with different shear geometries

The shear stress of ER fluids sheared between two parallel plates has been widely observed to increases quickly at small shear strains until it reaches the shear yield stress. But no shear stress jump phenomenon has been reported yet. Considering the difference of shearing ER fluids between two concentric cylinders and between two parallel plates, the shear rate in the former geometry is uniform, while in the later geometry, the shear rate increases linearly with the radius of rotation. So when shearing ER fluids between two parallel plates under high electric fields, the shear thickening is generally expected to happen gradually from the outmost radius to the inner radius while the average shear rate increases. So, there might be no abrupt shear stress jump when an ER fluid is sheared between two parallel plates. Also, in this geometry, ER fluids have been widely observed to be pushed out of the plate gap at high shear rates.[27-28] The inner chains have a tendency to move to a outer radius position. This may cause a forced lateral chain aggregation by shearing ER chains between two parallel plates, and result in stronger chain structures and higher shear strength. This effect may be more significant than the shear thickening happened between two concentric cylinders. Test an ER fluid with both geometries have been done. The ER fluid has a zero field viscosity of about 4 Pa.s at room temperature. The rheometer with concentric cylinder is still MCR 301. The rheometer with two parallel plates is HAAKE RV20 [15] with a plate diameter of 20 mm and a gap distance of 0.7 mm, the same as the gap in the cylindrical geometry. The shear rate linearly increased from 0 to 6 s$^{-1}$ in 120 seconds in the plate test. The ER fluid on the edge of the plate has been carefully trimmed off. Results are shown in Figure 6. As expected, the cylindrical geometry gave a lower shear stress but with obvious shear stress jumps. This effect may have implications for the design of ER actuators.

## 3.4 A proposed frictional mechanism for ER effect

The increase in viscosity of colloidal suspensions by two orders of magnitude even with a particle interaction potential $U \to 0$ during shearing thickening has been attributed to the mechanical contacts and the friction forces between particles.[4,5,24,25] On the other hand, for over 60 years, the shear yield stress of ER fluids has been mainly attributed to the attractive electrostatic interaction between particles.[7-11,14-15,26] However, there is no direct experimental quantification of the electrostatic interactions between real ER particles. Considering the aggregation of particles from single chains to columns in ER fluids, the Madelung constant for the many-body interaction of a crystal structure of usually considering a cubic or body-centered tetragonal structure)[14] could not account for the ratio of shear stress change during shear thickening, which could be as high as 9 or even higher under higher electric fields and particle volume fractions. Therefore, the direct mechanical contacts and friction forces between particles in a shear thickened state should also be the real origin of the high shear yield stress of ER fluid (sketched in Fig. 1c). In fact, recently, different from the traditional electrostatic/magnetic interaction between particles, the importance of friction force to the shear yield stress of ER and MR fluids prepared from fibers[29-31] and sea-urchin-like hierarchical morphology particles[32] has been discussed. For instance, column, zigzag, three-dimensional stochastic and near-planar stochastic structures of the magnetic fiber suspensions have been tested, the yield stress is found to mostly come from the restoring magnetic torque acting on each fiber and the solid friction between fibers[29-30]. In the case of spherical particles, as sketched in Fig. 1c, the particles are pushed together by the shear thickening-induced load $L$ and the attractive electrostatic force $F(r, \theta)$ between particles, which produce a friction force $F_f = f \cdot (F(r, \theta) + L)$, where $f$ is the friction coefficient. The electric field $E$ could increase the local particle concentration $\phi$, induce an additional internal load $F(r,$ θ), and a load *L* between the particles to facilitate the occurrence of shear thickening. *E* also controls the extent of shear thickening.

In the traditional polarization model, the particle interaction strength increases with the increase of the applied electric field. Since the particles only have perpendicular interaction with the electrode surface, there is generally a problem of generating resistance at the interface of ER fluid /electrodes. So generally, people thought the yielding of ER fluid does not occur at the ER fluid/electrode interface. However, with friction forces between particles and the particle/electrode interface, the shear resistance could be steadily transmitted from the electrode to the inner bulk materials of ER fluids. The applied external electric field could increase both the shear thickening induced mechanical repulsive force *L* and the particle interaction strength F(r, θ) to increase the resulted shear yield stress under electric fields. Thus, the shear yield stress based on the friction force between particles also increases with the increase of the external electric field. Many traditional descriptions of shear yield stress by electric field could be easily modified to the friction force description.

### 3.5 Discussions with previous experimental results

Considering the similarities between the ER effect and the MR effect, the friction force during shear thickening should also be directly responsible for the high shear yield stress of MR fluids. The friction force can explain why a normal compression of an ER/MR fluid can significantly increase the lateral shear yield stress of the ER/MR fluid by over one order of magnitude. This can also explain why the increase in yield stress is proportional to the applied pressure with a factor of about 0.3,[33] which is a typical value for the friction coefficient between two solid materials. The compression has a direct contribution to the load *L*. The shear yield stress $\tau_E$ of the ER fluid is related to the tensile yield stress $\sigma_E$ through $\tau_E = 0.3\sigma_E$.[34] Meanwhile, the shear thickening of a colloidal suspension consisting of smaller particles corresponds to a higher critical shear stress.[2] A similar inverse relationship between yield stress and particle diameter $a$ of $\tau_s \propto 1/a$ has been obtained in a giant ER fluid.[11] Increasing the particle diameter from 20 nm to 1 μm, the yield stress of the giant ER fluid[11] under 5 kV/mm should decrease from 250 to 5 kPa, the same level of general ER fluids composed of micron-sized particles.[14,15] Therefore, the giant ER effect is essentially the same as the general ER effect. Both of them should be electric field-controlled shear thickening. The local electric field between particles in a giant ER fluid does not need to be over 100 k/mm as calculated in previous reports.[10,16] This electric field is far above the breakdown electric field strength of most insulating liquids. Decreasing particle size and increasing the friction coefficient between particles should be effective ways to increase the shear yield stress of both ER and MR fluids.

The shear thickening in ER fluids, a kind of dipolar suspension, has implications for the shearing of other dipolar suspensions. For instance, "frozen" water confined within a nanometer gap and under a high electric field[35] may also undergo a similar shear thickening process and behave like a solid. A thin polymer melt film confined between two smooth mica surfaces can also undergo such a similar shear thickening process.[36] The above results may predict clutch fluids with high friction forces triggered by applying a local electric field.

Aside from electric/magnetic fields that could adjust particle interaction and induce shear thickening, other external stimuli such as light, heat, and electrochemistry may also be utilized to develop new field-responsive intelligent shear thickening materials.

## 4. Conclusions

Reversible electric field controlled shear thickening of ER fluids above certain low critical shear rate and high critical electric field strength has been found and verified in this study by shearing ER fluids between two concentric cylinders with various shearing modes. The onset of shear thickening in ER fluids could be characterized by a modified Mason number. While traditional theory generally ascribes the shear strength electrorheological (ER) fluids directly to the attractive electrostatic interaction between ER particles, the experimental results in this study shows that shear thickening-induced friction force between particles should be the real origin of the high shear strength of ER fluids. The applied external electric field can control the extent of shear thickening and affect the final shear strength of ER fluids. This electric field controlled shear thickening has implications for preparing advanced ER/MR fluids and studying the shear behaviors of other dipolar suspensions or liquids.

**Acknowledgments**

This work is supported by the National Natural Science Foundation of China (Grant No. 50875152) and the Program for New Century Excellent Talents in University of China. N.P., H.Z. and J.I. were supported by the U.S. Department of Energy, Office of Basic Sciences, Division of Materials Sciences and Engineering under Award # DE-FG02-87ER45331.

# Figures

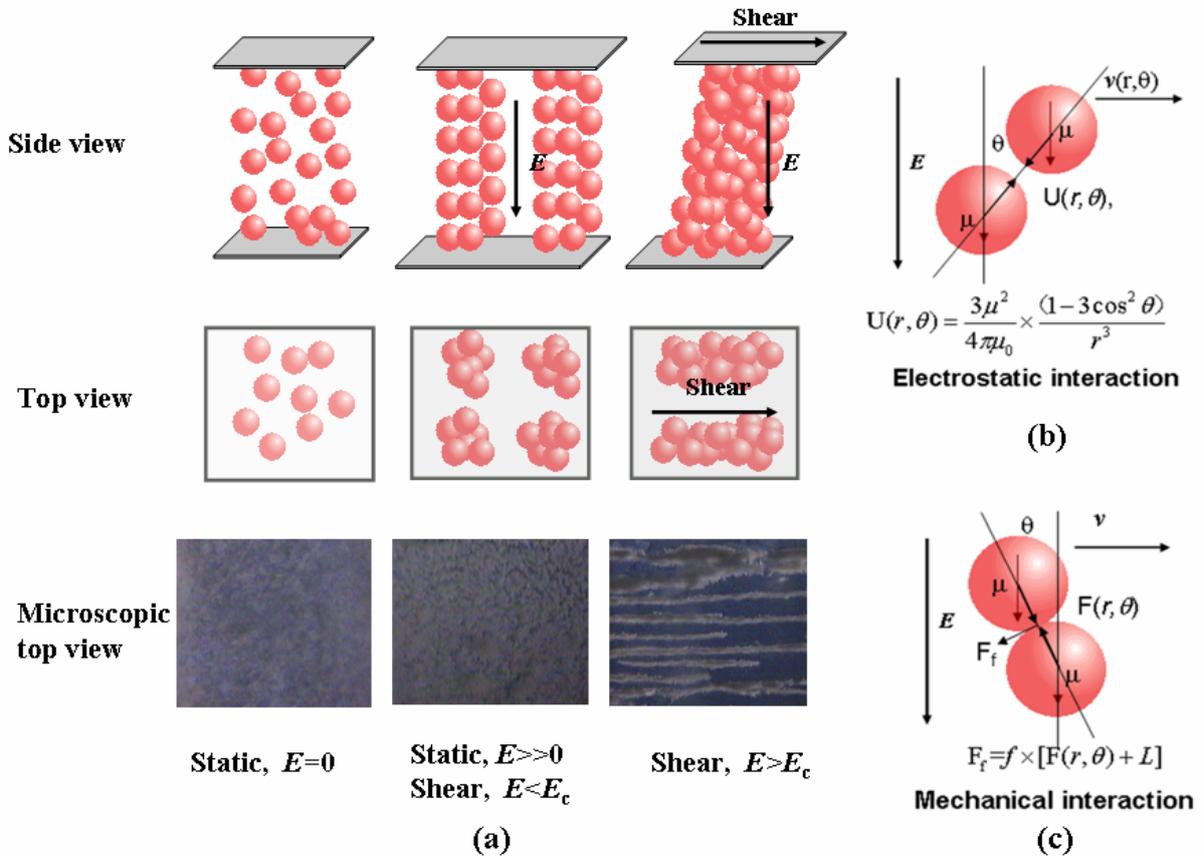

Figure 1. Sketches of the ER effect and the origin of its shear resistance. (a) Side view, top view, and microscope top view of ER fluids under different electric fields $E$ and shear states $\dot{\gamma}$. (b) Electrostatic pair interaction between particles in the traditional polarization model. (c) Mechanical interaction induced by the electrostatic interaction between particles after these particles become shear thickened or jammed.

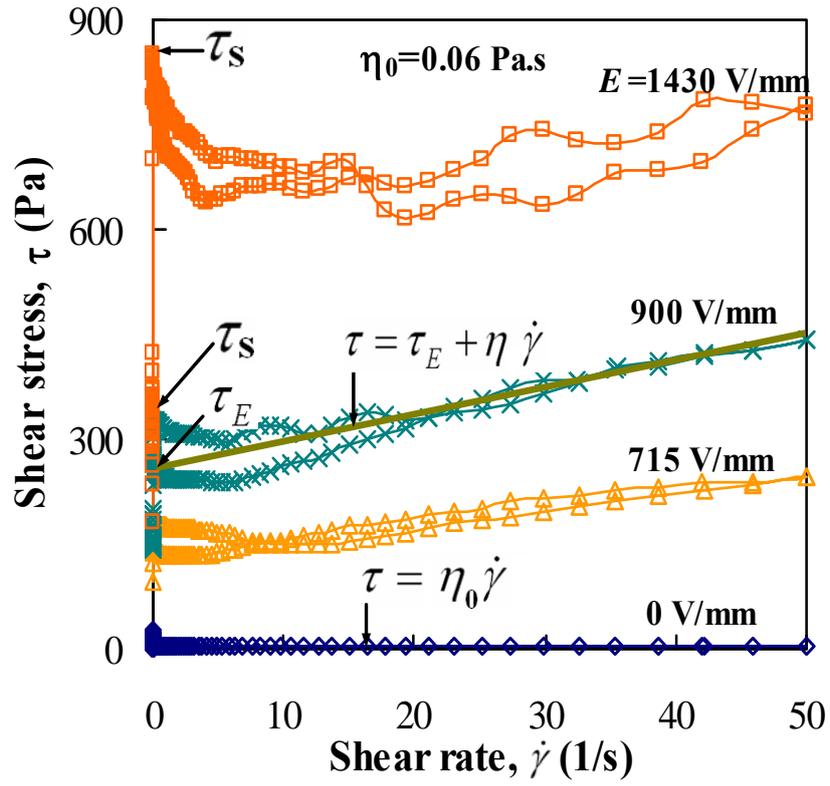

(a)

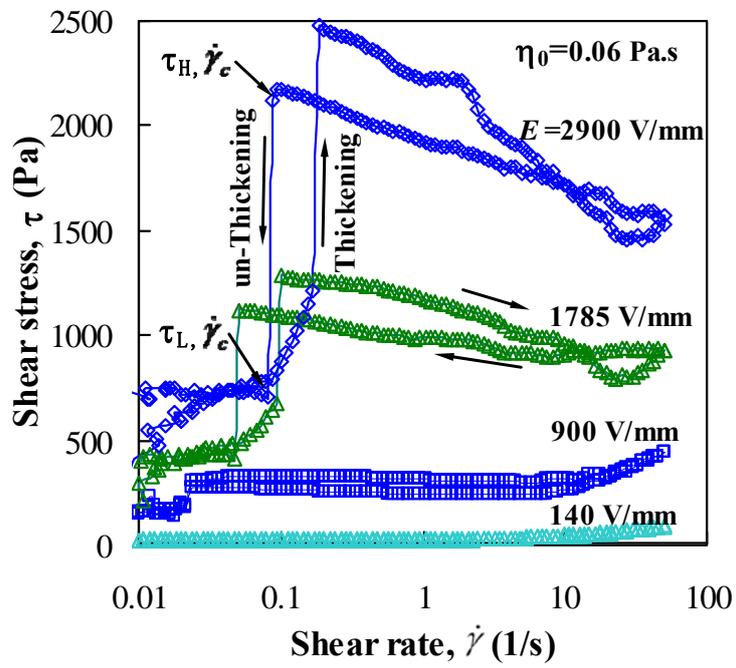

(b)

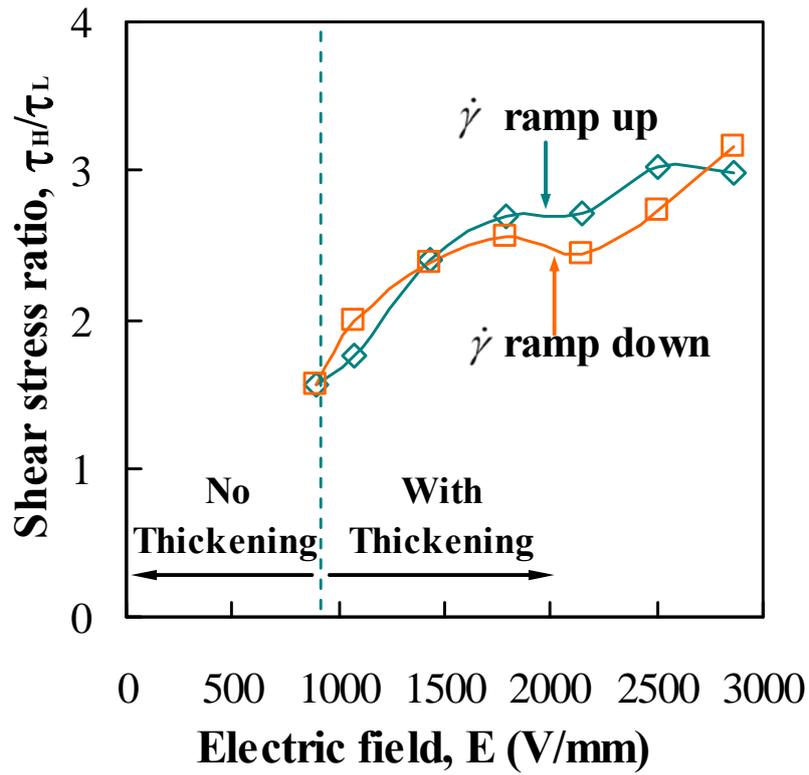

(c)

Figure 2. Shear test results of an ER fluid with a particle volume fraction of 23% on a commercial rheometer (Physica MCR 301). (a) Shear curves of the ER fluid at different electric fields. (b) Shear curves with logarithmically plotted shear rates. (c) Ratio of high to low shear stress during shear rate ramp up and ramp down under different electric fields.

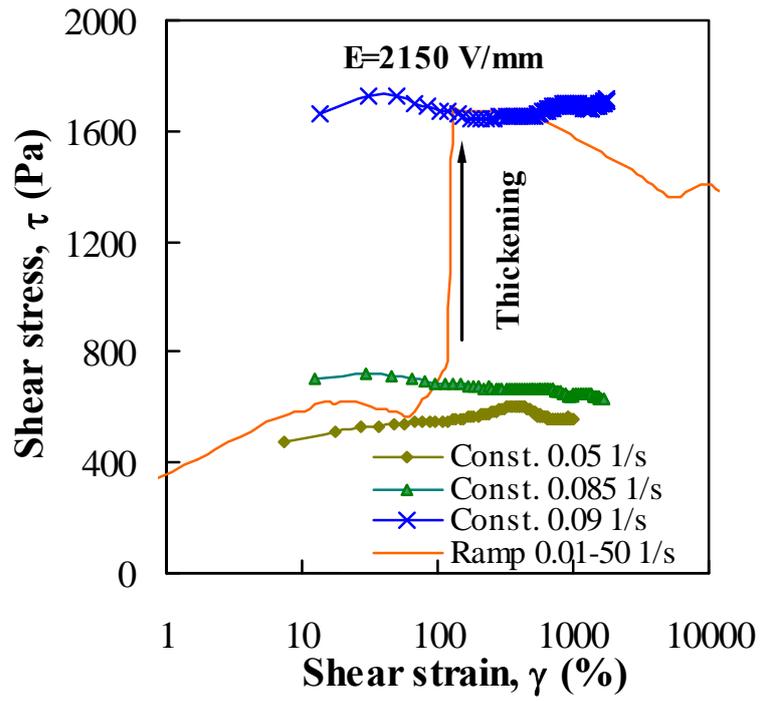

(a)

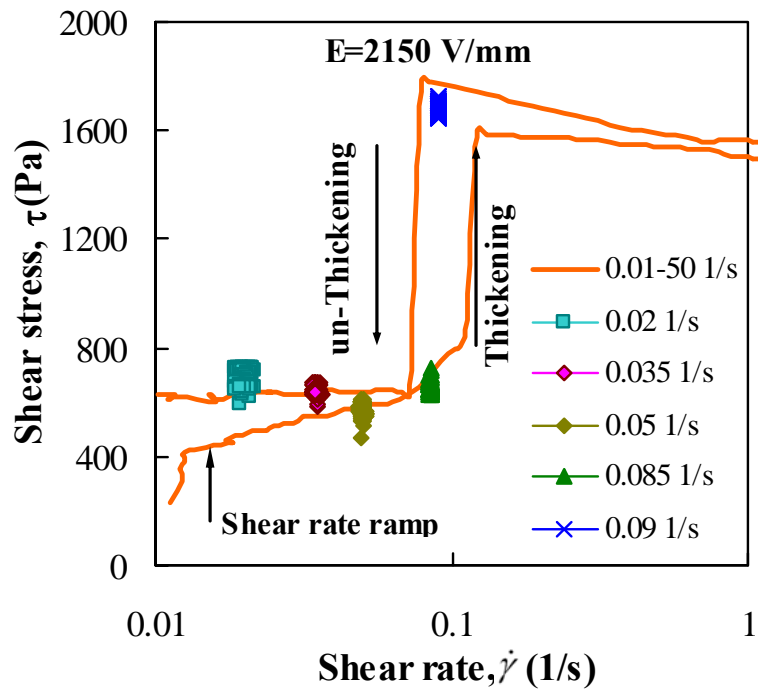

(b)

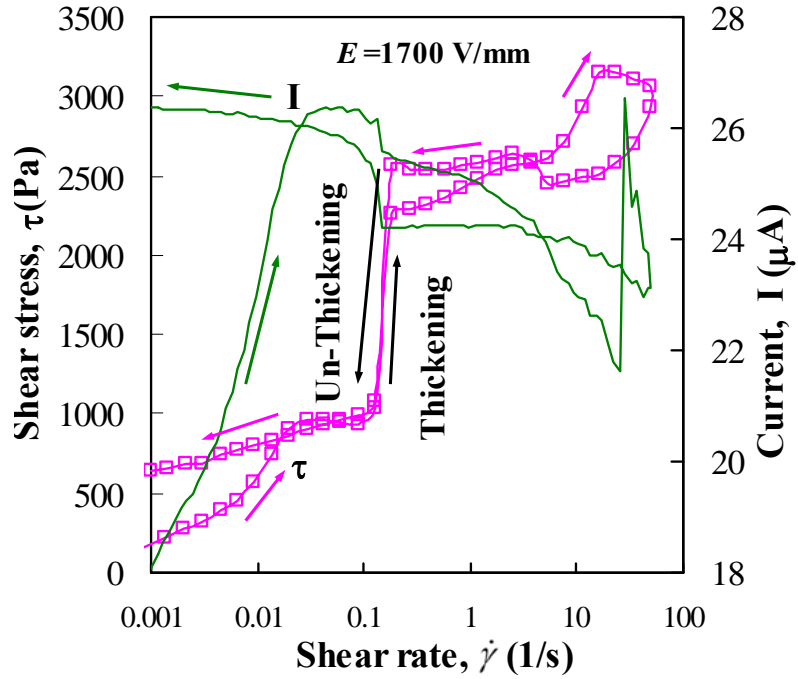

(c)

Figure 3. Shearing at different constant shear rates and a current measurement. (a) Shear stress versus shear strain at different constant shear rates or shear rate ramp up under the same electric field of 2150 V/mm. (b) Shear stress versus shear rate at different constant shear rates or shear rate ramp up under the same electric field of 2150 V/mm. (c) Typical results of shear stress and current of the ER fluid (particle volume fraction $\phi = 28\%$) in a shear rate ramp test.

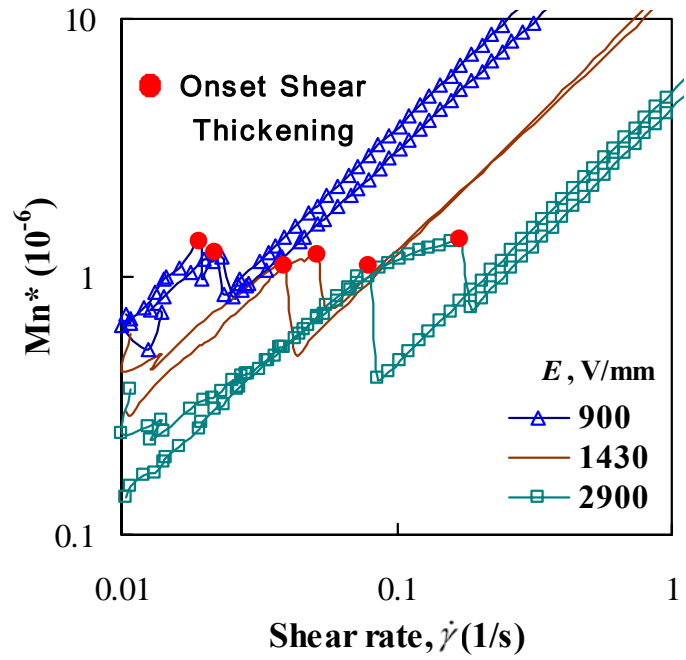

(a)

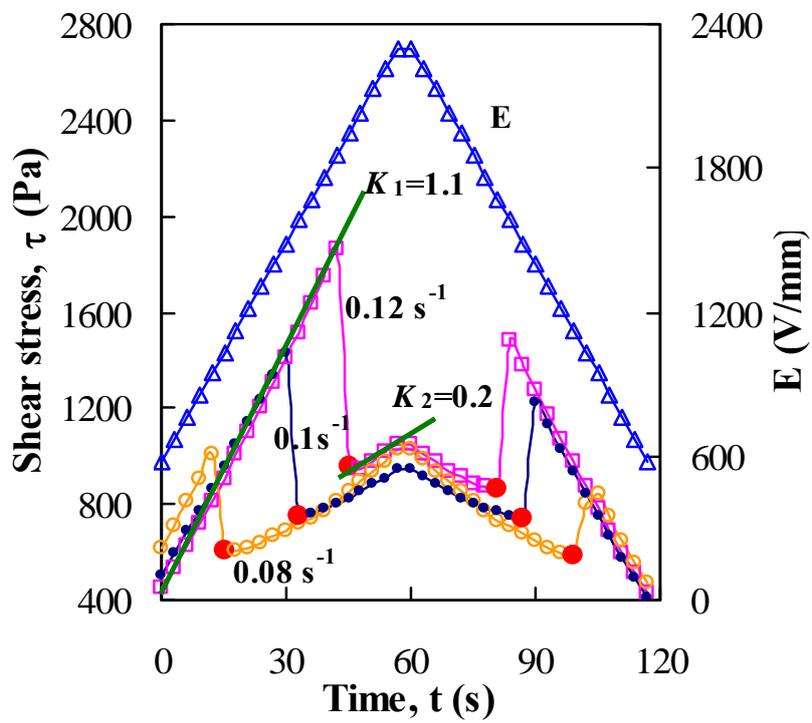

(b)

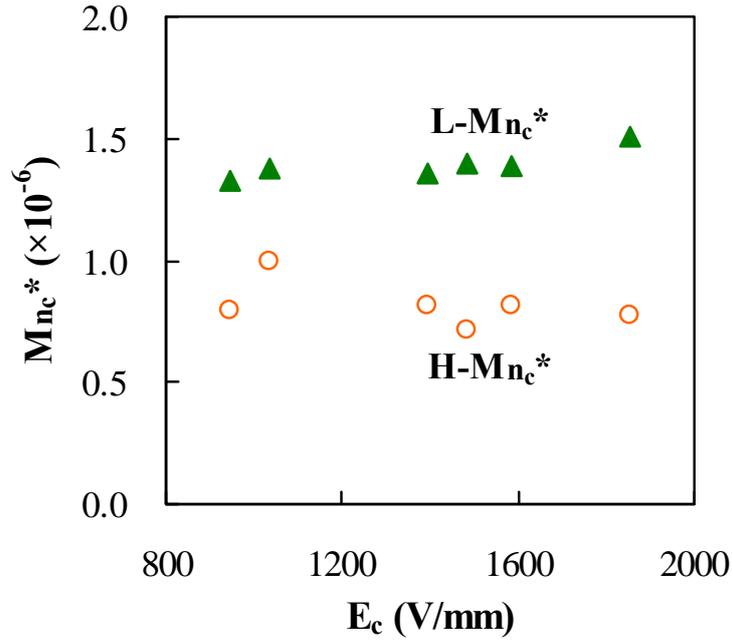

(c)

Figure 4. The modified Mason number $Mn^*$ in the shear thickening of ER fluids. (a) $Mn^*$ curves of shear rate ramp tests replotted from Fig. 2. (b) Electric field ramp tests at a constant shear rate of 0.08, 0.1, or 0.12 s$^{-1}$ with $E$ ramped between 580 and 2300 V/mm. (c) Critical Mason numbers $Mn_c^*$ for the shear thickening shown in Fig. 4(b).

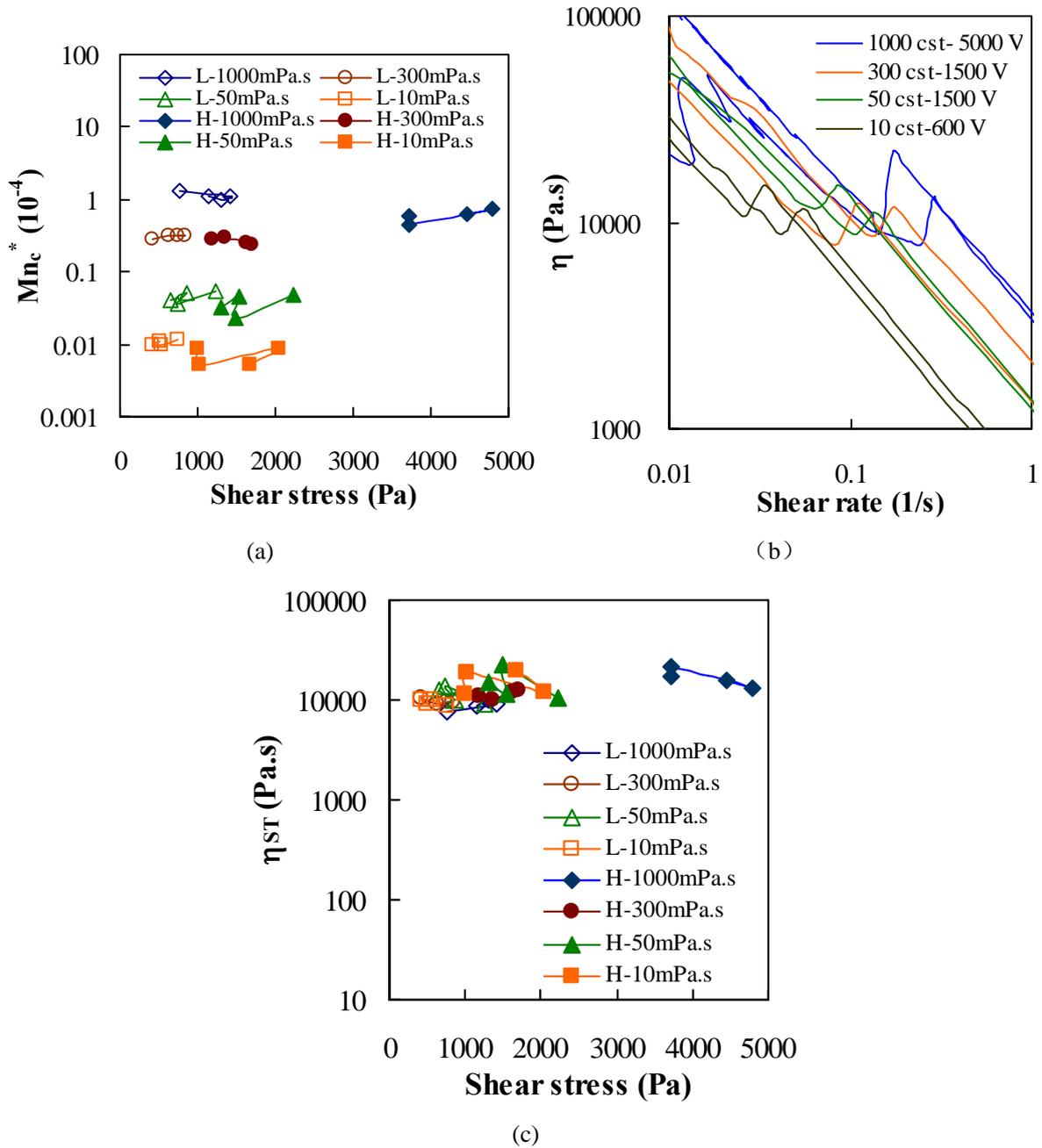

Figure 5 Results of ER fluids with different silicone oil viscosities of 1000 mPa.s, 300 mPa.s, 50 mPa.s and 10 mPa.s. (a) The critical Modified Mason numbers of ER fluids for shear thickening applied different electric fields, L means the low shear stress point before the shear stress jump, H means the high shear stress point after the shear stress jump; (b) Typical curves of the apparent viscosity of ER fluids during shearing; (c) The critical apparent viscosity of ER fluids for the shear thickening.

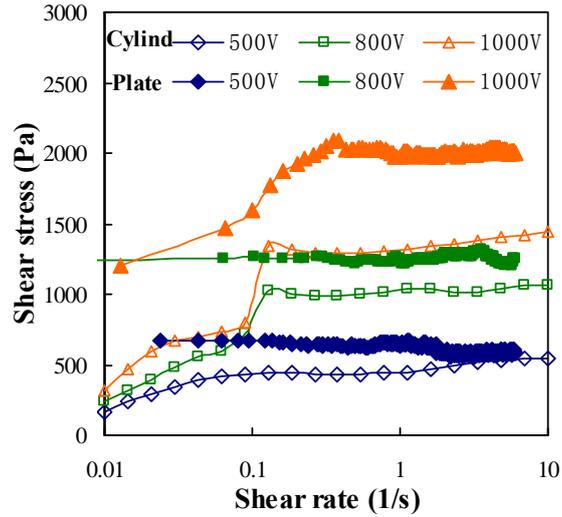

Figure 6 Comparison of shearing ER fluid between two concentric cylinders and between two parallel plates. The empty symbols are for the tests with two concentric cylinders. The filled symbols are for the tests with two parallel plates. During tests, the ER fluid at the edge of the plate has been carefully removed.

# Supporting Information

1. Shearing time effect on shear thickening effect

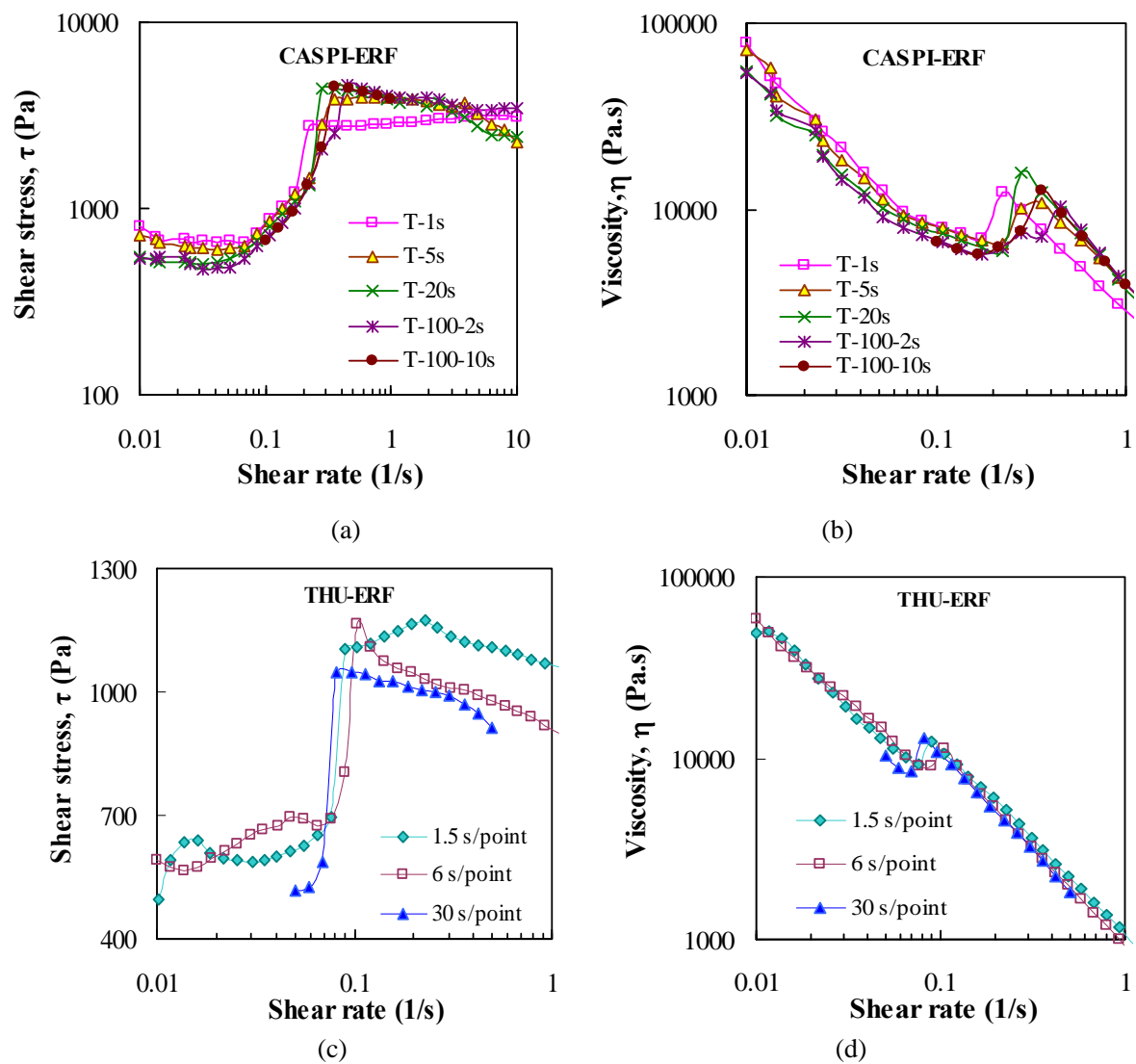

Fig. 1 Shearing time effect on the shearing curve. (a) Test curves of shear stress versus shear rate of CASPI-ERF under 5000 V; (b) curves of viscosity versus shear rate of CASPI-ERF under 5000 V; (c) Test curves of shear stress versus shear rate of THUI-ERF under 1200 V; (d) curves of viscosity versus shear rate of THU-ERF under 1200 V.

To test the shearing time effect on the shear thickening effect, two ER fluids were used. One is obtained from the Institute of Physics of Chinese Academy of Sciences (called CASPI-ERF, based on modified $TiO_2$ nano-particles and silicone oil with a viscosity of 50 mPa.s, the viscosity of CASPI-ERF at zero field is about 20 Pa.s at room temperature 20°C). The other is prepared by us, called THU-ERF. It is based on zeolite and silicone oil, viscosity at zero field is about 1.5 Pa.s at room temperature 20°C.

For CASPI-ERF, similar to our former experiment, shear rate is ramped up in the range of 0.01 to 10 s$^{-1}$ with different shearing time of 1 to 20 s at each shear rate, a flexible control of changing shearing time of 100 s at 0.01 s$^{-1}$ to 2 s at 10 s$^{-1}$ to promise a shear strain of higher than 1 at each shear rate, and a flexible control of changing shearing time of 100 s at 0.1 s-1 to 10 s at 1 to promise a shear strain of 10 at each shear rate. Results are shown in Fig. 1(a) and (b). The figures show that except the amplitude of the shear stress with a time of 1 s at each shear rate is a lower than in other tests, the shearing curves are very similar to each other. The points of shear thickening also reasonably agree with each other. In this test, the shear stress of the ER fluid sheared with an interval time of 1 s is lower than that with 5s or other test parameters. We prefer to take this as a shear history effect, rather than a shearing time effect due to the following tests.

For THU-ERF, the shear rate is ramped up in the range of 0.01 to 10 s$^{-1}$ with different shearing time of 1.5 to 6 s at each shear rate, or in the range of 0.05 to 0.5 with 30 s at each shear rate. The three curves agree with each other well as shown in Fig. 1(c) and (d). The amplitudes of the shear stresses are also close to each other. So it is not meaning a shorter shearing time corresponds to a lower shear stress. The amplitude of shear stress largely depends on the shear history.

2. Particle volume fraction effect on Shear thickening of ER fluids

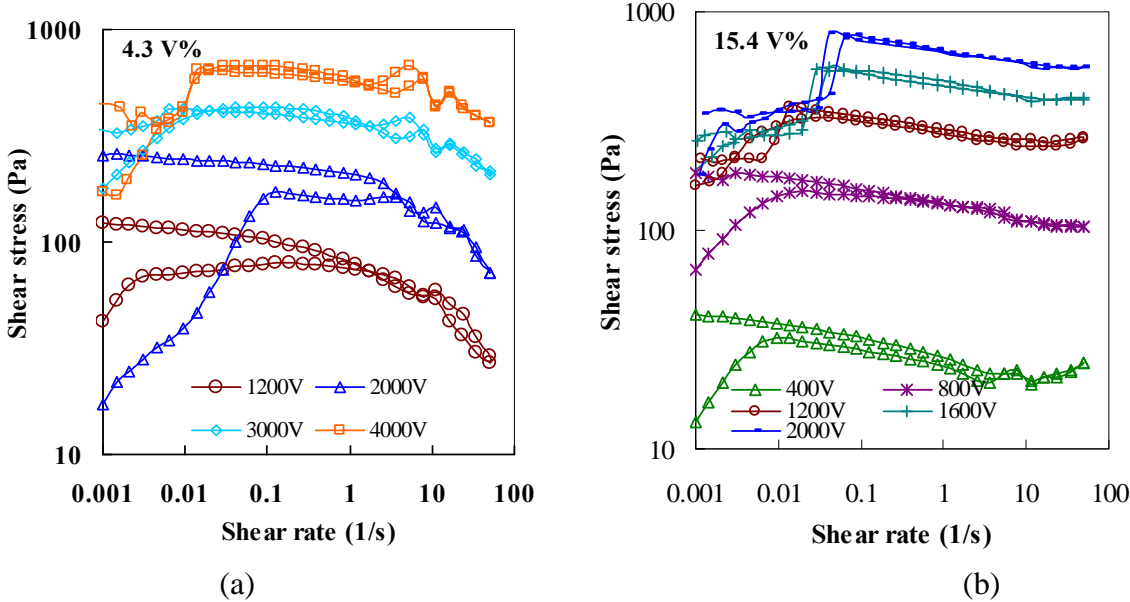

(a)  (b)

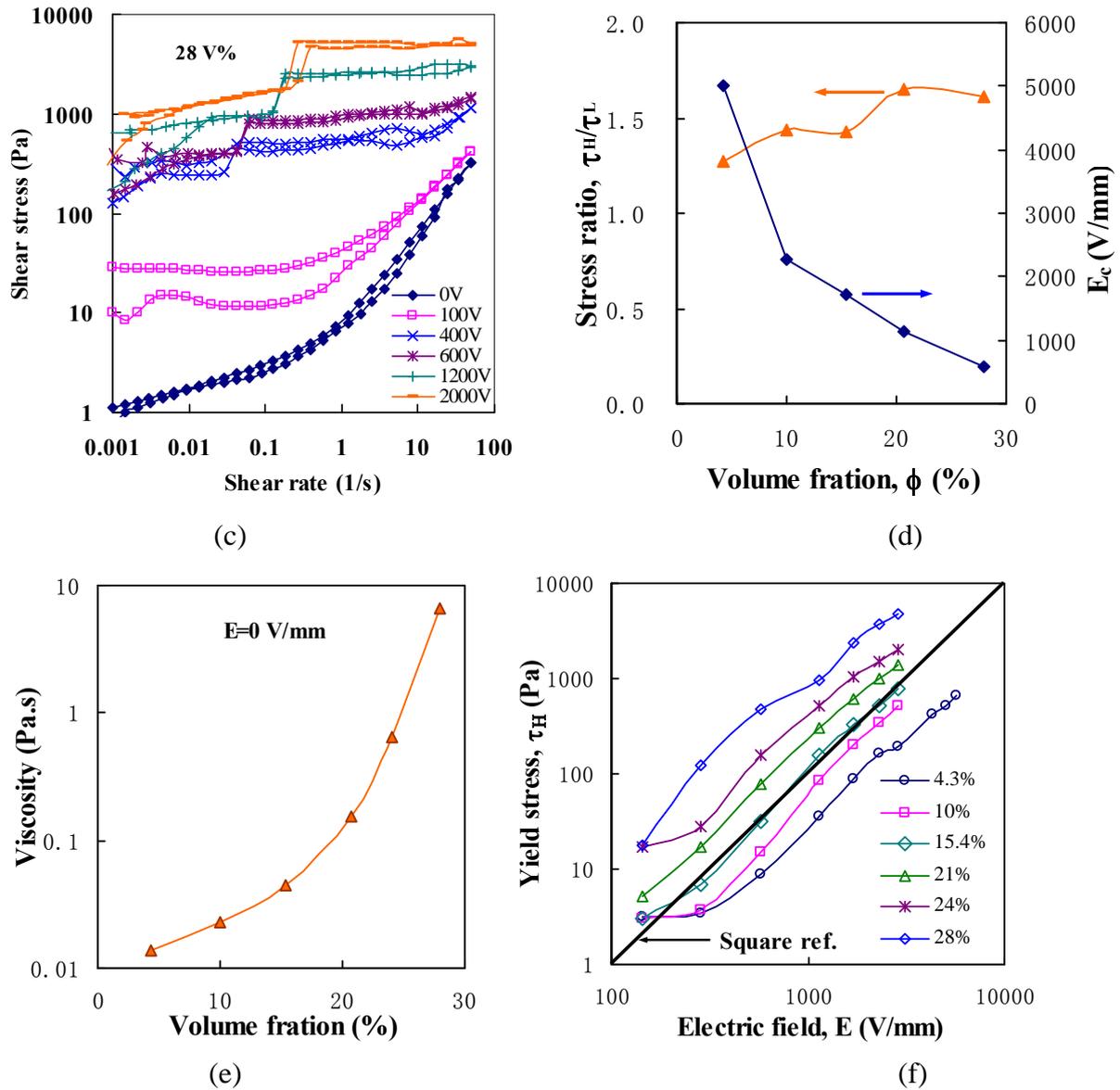

Fig. 2 Results of experimental data of ER fluid prepared with zeolite particles and silicone oil with different viscosities. (a) Particle volume fraction 4.3%; (b) Particle volume fraction 15.4%; (c) Particle volume fraction 28%; (d) the cirtical electric field for shear thickening to happen at different particle volume fractions; (e) the zero field viscosity of ER fluid with different particle volume fractions; (f) the yield stress of ER fluids with different particle volume fractions and electric fields.

ER fluids prepared with our generally used zeolite particles and silicone oil with a viscosity 50 mPa.s and with different particle volume fractions ranges from 4.3% to 28% have been tested using two concentric cylinders on MCR 301. Typical results are shown in Fig. 2 (a-c). l As shown in Fig. 2 (d), the critical electric field $E_c$ for shear thickening to happen decreases from 4667 V/mm (4.3 %) to 533 V/mm (28 %). The ratio of $\tau_H/\tau_L$ at those points increases with the increase of $\phi$ and $E$. The ratio is above 3 for $\phi$=28 % and $E$= 2900 V/mm. The zero field viscosity of the ER fluids increases quickly with the increase of the particle volume fraction as shown in Fig. 2(e). Taking $\tau_H$ as the shear yield stress, the relationship between $\tau_H$ and *E* roughly agrees with the square relationship predicted by traditional polarization models as shown in Fig. 2(f).

3. Medium viscosity effect on Shear thickening of ER fluids

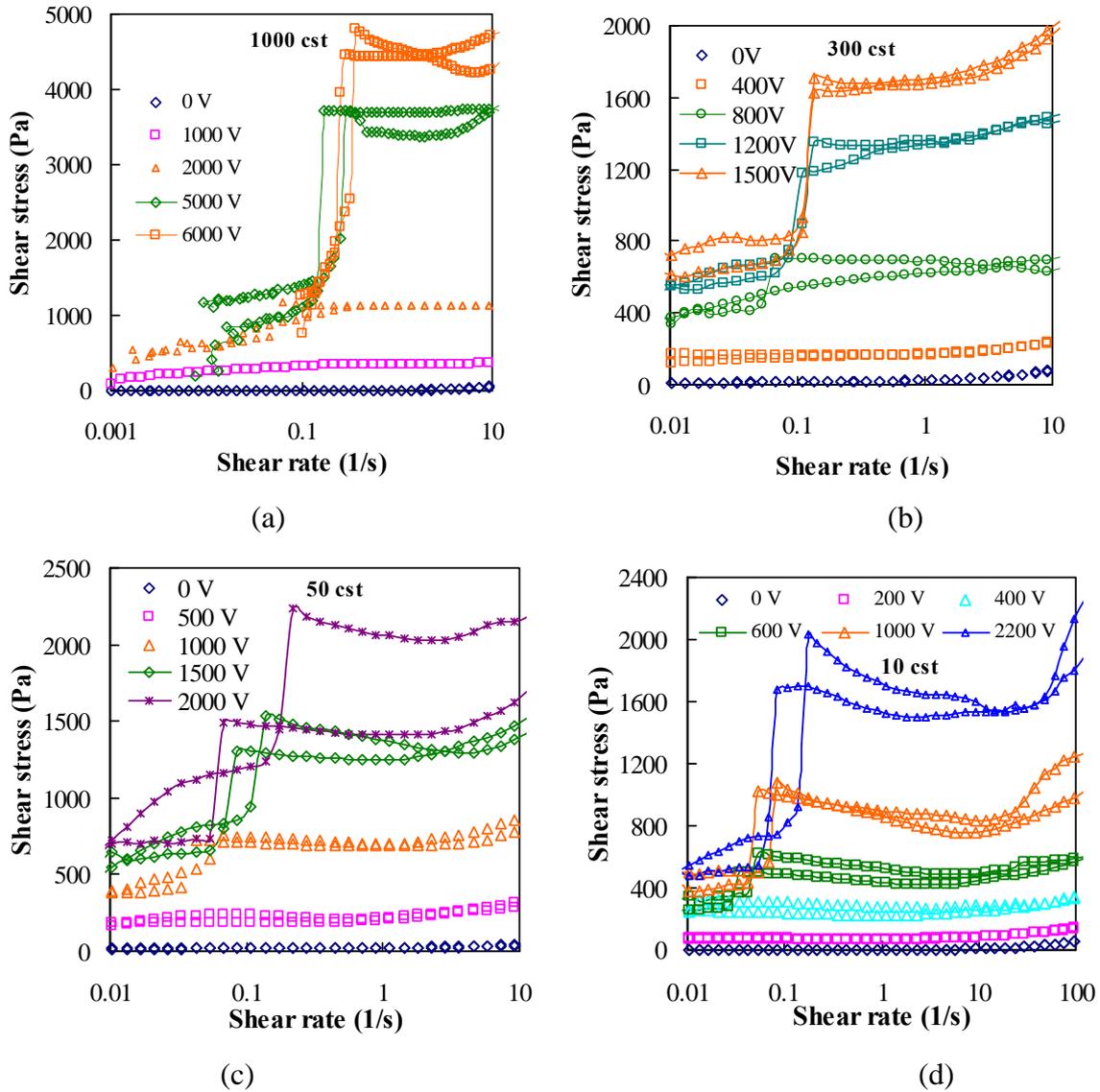

Fig. 3 The original data of shearing ER fluids prepared with different insulating liquid viscosities. (a) The ER fluid using silicone oil with a viscosity of about 1000 mPa.s; (b) The ER fluid using silicone oil with a viscosity of about 300 mPa.s; (c) The ER fluid using silicone oil with a viscosity of about 50 mPa.s; (d) The ER fluid using silicone oil with a viscosity of about 10 mPa.s.

All of the ER fluids are prepared based on our generally used zeolite particles and silicone oil. The viscosity of the silicone oil is at room temperature 20°C.